# ROBUST STABILIZATION OF LINEAR PLANTS UNDER UNCERTAINTIES AND HIGH-FREQUENCY MEASUREMENT NOISES[1]

Igor B. Furtat*,**. Artem N. Nekhoroshikh, *,**

*Institute for Problems of Mechanical Engineering Russian Academy of Sciences, 61 Bolshoy ave V.O., St.-Petersburg, 199178, Russia (Tel: +7-812-321-47-66; e-mail: cainenash@mail.ru).
**ITMO University, 49 Kronverkskiy ave, Saint Petersburg, 197101, Russia.

**Abstract:** The paper describes the robust algorithm for linear time-invariant plants under parametric uncertainties, external disturbances and high-frequency noises in measurements. The proposed algorithm allows one to reduce the noise impact on the output variable of the plant and to compensate parametric uncertainties and external disturbances independently. The modeling results illustrate the effectiveness of the algorithm.

## 1. INTRODUCTION

Design of simple control systems under parametric uncertainties and external disturbances when only plant output is available for measurement is important problem of control theory and practice. To construct such control schemes many solutions have been proposed in this regard.

One of the most effective tools is to synthesize control structure using high-gain observer. At first high-gain observer was proposed in (Esfandiary and Khalil, 1992; Gauthier et al., 1992) for minimum phase plants. Later other kind of high-gain observer were considered in (Gauthier et al., 1992; Teel and Praly, 1994; Bobtsov, 2002; Tsykunov, 2008; Furtat, 2015). In (Esfandiary and Khalil, 1992; Gauthier et al., 1992; Teel and Praly, 1994; Bobtsov, 2002; Furtat, 2015) the dimension of the high-gain observer is equal to $\gamma - 1$, where $\gamma$ is the relative degree of plant model.

However, using high-gain observer can be unsatisfactory in case high frequency noise measurement application. The investigations of the high-gain observers under noises were considered in (Vasiljevic and Khalil, 2008; Boizot et al., 2010; Sanfelice and Praly, 2011). The problem is that the value of estimate derivative could be sufficiently greater than the real one. Moreover, the error accumulates in further estimation.

In (Ahrens and Khalil, 2009; Sanfelice and Praly, 2011; Prasov and Khalil, 2013) adaptive high-gain observer was proposed to partially overcome this problem. Thus, initially high-gain parameter of the observer has a large value, while in steady state mode high-gain parameter is decreased.

In (Astolfi and Marconi, 2015; Wang et al., 2015) an extended high-gain observer was considered. The dimension of the modified observer is $2\gamma - 2$. Doubling dimension is caused by the introduction of additional differential equations reducing the impact of high frequency measurement noises. The simulation results showed the effectiveness of the modified algorithm as compared with the standard high-gain observer. However, in (Astolfi and Marconi, 2015; Wang et al., 2015) quality of estimate derivatives and quality of filtering simultaneously depend on the solution of the observer equation.

In the present paper we consider two independent algorithms: filtering and control ones. Differently from (Astolfi and Marconi, 2015; Wang et al., 2015), the proposed algorithm allows one

1) to improve the quality of the estimation derivatives;

2) to calculate independently the filter parameters and the parameters of the observer.

The paper is organized as follows. The problem statement is presented in Section 2. In Section 3 we design the high-frequency filtering algorithm. In Section 4 we propose the control algorithm for linear plants. In Section 5 we consider simulation results and discuss an efficiency of the proposed control structure. Concluding remarks are given in Section 6.

---

[1] The results of Section 3 was developed under support of RSF (grant 14-29-00142) in IPME RAS. The results of Section 4 wsa supported solely by the Russian Federation President Grant (No. 14.W01.16.6325-MD (MD-6325.2016.8)). The other research were partially supported by grants of Russian Foundation for Basic Research No. 16-08-00282, № 16-08-00686, Ministry of Education and Science of Russian Federation (Project 14.Z50.31.0031) and Government of Russian Federation, Grant 074-U01.

## 2. PROBLEM STATEMENT

Consider a plant model in the form

$$Q(p)z(t) = kR(p)u(t) + f(t), \quad y(t) = z(t) + w(t), \quad (1)$$

where $y(t) \in R$ is an output, $u(t) \in R$ is an input, $f(t) \in R$ is a unmeasured bounded disturbance, $w(t) \in R$ is a high frequency bounded noise, $Q(p)$, $R(p)$ are linear differential operators with unknown coefficients, $\deg Q(p) = n$, $\deg R(p) = m$, $k > 0$, $p = d/dt$.

Assume that the coefficients of operators $Q(p)$, $R(p)$ and coefficient $k > 0$ belong to a known compact set $\Xi$. The polynomial $R(\lambda)$ is Hurwitz, where $\lambda$ is a complex variable.

The problem is to design the control system such that the following condition holds

$$\overline{\lim_{t \to \infty}} |z(t)| < \delta_1, \quad (2)$$

where $\delta_1 > 0$ is a required accuracy, hereinafter $|\cdot|$ is an Euclidean norm.

## 3. HIGH FREQUENCY FILTERING ALGORITHM

Reject signal $w$ from signal $y$. To this end, introduce the following algorithm

$$\mu \dot{\xi}(t) = G\xi(t) + By(t), \quad \hat{y}(t) = L\xi(t), \quad \xi(0) = 0, \quad (3)$$

where $\xi = [\xi_1, \xi_2, ..., \xi_r]^T$,

$$G = \begin{bmatrix} -\sigma_1^{-1} & 0 & 0 & \cdots & 0 & 0 \\ \sigma_2^{-1} & -\sigma_2^{-1} & 0 & \cdots & 0 & 0 \\ 0 & \sigma_3^{-1} & -\sigma_3^{-1} & \cdots & 0 & 0 \\ \vdots & \vdots & \vdots & \ddots & \vdots & \vdots \\ 0 & 0 & 0 & \cdots & \sigma_r^{-1} & -\sigma_r^{-1} \end{bmatrix}, \quad \sigma_i > 0,$$

$B = [\sigma_1^{-1}, 0, ..., 0]^T$, $\mu > 0$ is a sufficient small coefficient, $L = [0, ..., 0, 1]$.

*Theorem 1.* Let signal $z$ be bounded and the following condition holds

$$\mu^{-1} \overline{\lim_{t \to \infty}} \left| \int_0^t e^{\mu^{-1}G(t-s)} Bw(s)ds \right| < \delta_2, \quad (4)$$

where $\delta_2 > 0$ is a sufficiently small coefficient. Then there exists a coefficient $\mu_0 > 0$ such that for any $\mu \le \mu_0$ the following condition holds

$$\overline{\lim_{t \to \infty}} |\hat{y}(t) - z(t)| < \delta_3. \quad (5)$$

Here $\delta_3 > 0$ is a sufficiently small coefficient.

*Proof of Theorem 1.* Consider plant (3) with input signal $z$:

$$\mu \dot{\tilde{\xi}}(t) = G\tilde{\xi}(t) + Bz(t), \quad \tilde{y}(t) = L\tilde{\xi}(t), \quad \tilde{\xi}(0) = 0. \quad (6)$$

For analysis of plant (6) let us use Lemma (Furtat, 2014; Furtat et al., 2014; Furtat et al., 2015).

Lemma. *Consider a plant model*

$$h\dot{x}(t) = f(x(t), u(t)), \quad (7)$$

*where $x \in R^s$, $f(x, u, h)$ is Lipschitz function in $x$ and $u$, $u$ is a bounded function, $h > 0$ is a small coefficient. Let system (7) be asymptotically stable when $u = 0$. Consider the set $\Omega = \{x : f(x, u) = 0\}$. Then there exists $h_0 > 0$ for any $\varepsilon > 0$ such that for any $h < h_0$ the following condition holds*

$$\overline{\lim_{t \to \infty}} \mathrm{dist}(x(t), \Omega) < \varepsilon.$$

Let us verify conditions of Lemma for system (6). System (6) is asymptotically stable for $z = 0$, because the matrix $G$ is Hurwitz. Substituting $\mu = 0$ into (6), we get $G\xi = -Bz$ or $\xi_1 = z$ and $\xi_i = \xi_{i+1}$, $i = \overline{2, r-1}$. Thus, $\tilde{y} = z$. According to Lemma, there exists $\mu > 0$ such that for any $\mu \le \mu_0$ the following condition holds

$$\overline{\lim_{t \to \infty}} |\tilde{y}(t) - z(t)| < \delta_4, \quad (8)$$

where $\delta_4 > 0$ is a sufficiently small coefficient.

Consider the signal $y$ consisting of the signal $z$ and the noise $w$. Find a condition such that (5) will be hold.

Taking into account (3) and (6), rewrite the error $\zeta = \xi - \tilde{\xi}$ in the following form

$$\dot{\zeta}(t) = \frac{1}{\mu}G\zeta(t) + \frac{1}{\mu}Bw(t), \quad \hat{y}(t) - \tilde{y}(t) = L\zeta(t). \quad (9)$$

The solution of the first equation of (9) is

$$\zeta(t) = e^{\mu^{-1}Gt}\zeta(0) + \mu^{-1}\int_0^t e^{\mu^{-1}G(t-s)} Bw(s)ds$$
$$= \mu^{-1}\int_0^t e^{\mu^{-1}G(t-s)} Bw(s)ds. \quad (10)$$

If condition (4) holds, then we have from (10) that

$$\overline{\lim_{t \to \infty}} |\zeta(t)| < \delta_2. \quad (11)$$

Consider the following relations

$$|\hat{y}(t) - z(t)| \le |\hat{y}(t) - \tilde{y}(t)| + |\tilde{y}(t) - z(t)|$$
$$\le |\zeta(t)| + |\tilde{y}(t) - z(t)|. \quad (12)$$

Obviously, the upper bounds of (12) will be satisfied

$$\overline{\lim_{t \to \infty}} |\hat{y}(t) - z(t)| \le \overline{\lim_{t \to \infty}} |\zeta(t)| + \overline{\lim_{t \to \infty}} |\tilde{y}(t) - z(t)|. \quad (13)$$

Let $\delta_2 + \delta_4 \le \delta_3$. Taking into account (8) and (11), we get estimate (5) from (13). Theorem 1 is proved.

Let noise $w$ be sinusoidal signals

$$w(t) = \sum_{i=1}^{v} A_i \sin(\omega_i t + \varphi_i), \quad (14)$$

where $A_i$, $\omega_i$ and $\varphi_i$ are the amplitude, the frequency and the phase accordingly.

*Theorem 2.* Let z be bounded function and noise w be signal (14). Then there exists $\mu_0 > 0$ such that for any $\mu \leq \mu_0$ the following condition holds

$$\overline{\lim_{t\to\infty}}|\hat{y}(t) - z(t)| \leq \sum_{i=1}^{v} A_i \left( \prod_{j=1}^{r} \frac{1}{\sqrt{1+\omega_i^2\mu^2\sigma_j^2}} \right) + \delta_4. \quad (15)$$

*Proof of Theorem 2.* It follows from the proof of Theorem 1, that there exists $\mu > 0$ for system (6) such that for any $\mu \leq \mu_0$ the condition (8) holds. Taking into account (3) and (6), write the error $\zeta = \xi - \tilde{\xi}$ in the form (9). Rewrite system (9) as

$$\dot{\zeta}_1(t) = -\mu^{-1}\sigma_1^{-1}\zeta_1(t) + \mu^{-1}\sigma_1^{-1}\sum_{i=1}^{v} A_i \sin(\omega_i t + \varphi_i),$$
$$\dot{\zeta}_j(t) = -\mu^{-1}\sigma_j^{-1}\zeta_j(t) + \mu^{-1}\sigma_j^{-1}\zeta_{j-1}(t), \quad j = \overline{2, r}. \quad (16)$$

The solution of the first equation in (16) is

$$\zeta_1(t) = \mu^{-1}\sigma_1^{-1}\sum_{i=1}^{v} A_i \int_0^t e^{\mu^{-1}\sigma_1^{-1}(s-t)} \sin(\omega_i s + \varphi_i) ds$$
$$+ e^{-\mu^{-1}\sigma_1^{-1}t}\zeta_1(0) = \sum_{i=1}^{v} \frac{A_i}{\sqrt{1+\omega_i^2\mu^2\sigma_1^2}} \sin(\omega_i t + \varphi_i + \vartheta_{1,i}), \quad (17)$$

$$\vartheta_{1,i} = \arccos\frac{1}{\sqrt{1+\omega_i^2\mu^2\sigma_1^2}}.$$

Substituting (17) into the second equation of (16), we get

$$\zeta_2(t) = e^{-\mu^{-1}\sigma_2^{-1}t}\zeta_2(0)$$
$$+ \mu^{-1}\sigma_2^{-1}\sum_{i=1}^{v} \frac{A_i}{\sqrt{1+\omega_i^2\mu^2\sigma_1^2}} \int_0^t e^{\mu^{-1}\sigma_2^{-1}(s-t)}$$
$$\times \sin(\omega_i s + \varphi_i + \vartheta_{1,i}) ds$$
$$= \sum_{i=1}^{v} A_i \frac{1}{\sqrt{(1+\omega_i^2\mu^2\sigma_1^2)(1+\omega_i^2\mu^2\sigma_2^2)}}$$
$$\times \sin(\omega_i t + \varphi_i + \vartheta_{1,i} + \vartheta_{2,i}), \vartheta_{2,i} = \arccos\frac{1}{\sqrt{1+\omega_i^2\mu^2\sigma_2^2}}.$$

Therefore, the solution of the *r*th equation of (16) is

$$\zeta_r(t) = \sum_{i=1}^{v} A_i \left( \prod_{j=1}^{r} \frac{1}{\sqrt{1+\omega_i^2\mu^2\sigma_j^2}} \right) \sin(\omega_i t + \varphi_i + \sum_{j=1}^{r} \vartheta_{j,i}), \quad (18)$$

where $\vartheta_{j,i} = \arccos\frac{1}{\sqrt{1+\omega_i^2\mu^2\sigma_j^2}}$.

Upper bound of (18) is

$$|\zeta_r(t)| \leq \sum_{i=1}^{v} A_i \left( \prod_{j=1}^{r} \frac{1}{\sqrt{1+\omega_i^2\mu^2\sigma_j^2}} \right). \quad (19)$$

Takin into account (8), (13) and (19), we get estimate (15). Theorem 2 is proved.

## 4. SYNTHESIS OF CONTROL SYSTEM

Let us use the algorithm (Furtat, 2015) to synthesize the control system. According to Problem Statement, only the output signal y(t) is available for measurement. Therefore, introduce the control law as follows

$$u(t) = -\alpha \sum_{i=0}^{\gamma-1} d_i \overline{y}^{(i)}(t), \quad (20)$$

where $\alpha > 0$ and $d_0, d_1, ..., d_{\gamma-1}$ are chosen such that the polynomial $D(\lambda) = d_{\gamma-1}\lambda^{\gamma-1} + d_{\gamma-2}\lambda^{\gamma-2} + ... + d_1\lambda + d_0$ is Hurwitz, $\gamma = n - m$ is a relative degree of (1), $\overline{y}^{(i)}(t)$ is an estimate of the *i*th derivative signal $\hat{y}(t)$, $i = 0, 1, ..., \gamma - 1$.

Substituting (20) into (1), we get

$$F(p)z(t) = \alpha k R(p)g(t) + \alpha k R(p)D(p)\psi(t) + f(t) \quad (21)$$

where $F(p) = Q(p) + \alpha k R(p)D(p)$, $g(t) = D(p)\hat{y}(t) - \sum_{i=0}^{\gamma-1} d_i \overline{y}^{(i)}(t)$, $\psi(t) = z(t) - \hat{y}(t)$. The value of $g(t)$ depends on estimation quality of the signal $\hat{y}(t)$ and its derivatives, the value of $\psi(t)$ depends on quality of (3). Since the set $\Xi$ is known, then there exist the number $\alpha$ and the polynomial $D(\lambda)$ such that the polynomial $F(\lambda)$ is Hurwitz.

To implement the control law (20) we use the following observer

$$\overline{y}(t) = \hat{y}(t),$$
$$\overline{y}^{(j)}(t) = \frac{\overline{y}^{(j-1)}(t) - \overline{y}^{(j-1)}(t-h)}{h}, \quad j = \overline{1, \gamma-1}. \quad (22)$$

Substituting (22) into (20), rewrite the control law (20) in the form

$$u(t) = -\alpha \sum_{i=0}^{\gamma-1} \left[ \frac{d_i}{h^i} \sum_{j=0}^{i} (-1)^j C_i^j \hat{y}(t - jh) \right], \quad (23)$$

where $C_i^j = \frac{i!}{j!(i-j)!}$.

*Theorem 3.* Let w be a bounded signal. Additionally let ($\gamma - r - 1$)th derivatives of w be bounded, if $r \leq \gamma$. Then there exist coefficients $\alpha > 0$ and $h > 0$ such that the control system consisting of filtering algorithm (3) and control law (23) ensures goal (2).

*Proof of Theorem 3.* Transform system (21) to the form

$$\dot{\varepsilon}(t) = A\varepsilon(t) + \alpha k B_1 g(t) + \alpha k B_2 \psi(t) + B_3 f(t), \quad z(t) = J\varepsilon(t), \quad (24)$$

where $\varepsilon = [\varepsilon_1, \varepsilon_2, ..., \varepsilon_n]^T$, $z^{(i)} = \varepsilon_{i+1}$, $i = \overline{0, n-1}$, matrixes A, $B_1$, $B_2$, $B_3$ and $J = [1, 0, ..., 0]$ are obtained at the transition from (21) to (24). Rewrite system (3) as the following differential equation

$$\prod_{i=1}^{r} (\mu\sigma_i p + 1)\hat{y}(t) = y(t). \quad (25)$$

Transform (25) to the state space form

$$\dot{\theta}(t) = M\theta(t) + Ny(t), \quad \hat{y}(t) = J\theta(t), \qquad (26)$$

where $\theta = [\theta_1, \theta_2, ..., \theta_r]^T$, $\hat{y}^{(i)} = \theta_{i+1}$, $i = \overline{0, r-1}$, matrix $M$ and vector $N$ are obtained at the transition from (25) to (26).

Consider two cases.

1) Let $r < \gamma$. Rewrite the operator $D(p)$ in the following form

$$D(p) = \rho_1^T [p^{\gamma-1}, p^{\gamma-2}, ..., p^r] + \rho_2^T [p^{r-1}, p^{r-2}, ..., 1], \qquad (27)$$

where $\rho_1$ and $\rho_2$ are vectors with corresponding coefficients of the operator $D(p)$. Taking into account (27), rewrite function $g(t)$ in the form

$$g(t) = \sum_{j=r}^{\gamma-1} \rho_{1,j} J\theta^{(j)} + \rho_2^T \theta(t) - \sum_{i=0}^{\gamma-1} \frac{d_i}{h^i} J\theta(t) \\ - \sum_{i=1}^{\gamma-1} \left[ \frac{d_i}{h^i} \sum_{j=1}^{i} (-1)^j C_i^j J\theta(t - jh) \right], \qquad (28)$$

where $\rho_{1,j}$ is the $j$th element of the vector $\rho_1$. Taking into account (26), we find the $j$th derivative ($j \geq 1$) of $\theta$ in the following form

$$\theta^{(j)} = M^j \theta + \sum_{i=0}^{j-1} M^{j-i-1} Ny^{(i)} = M^j \theta + \sum_{i=0}^{j-1} M^{j-i-1} Nz^{(i)} \\ + \sum_{i=0}^{j-1} M^{j-i-1} Nw^{(i)} = M^j \theta + G_j \varepsilon + \sum_{i=0}^{j-1} M^{j-i-1} Nw^{(i)}, \qquad (29)$$

where $G_j = [N, MN, ..., M^{j-1}N, O, ..., O]$, $O$ is the zero matrix.

Substituting (29) into (28), we get

$$g = \sum_{j=r}^{\gamma-1} \rho_{1,j} J \left( M^j \theta + G_j \varepsilon + \sum_{i=0}^{j-1} M^{j-i-1} Nw^{(i)} \right) + \rho_2^T \theta(t) \\ - \sum_{i=0}^{\gamma-1} \frac{d_i}{h^i} J\theta(t) - \sum_{i=1}^{\gamma-1} \left[ \frac{d_i}{h^i} \sum_{j=1}^{i} (-1)^j C_i^j J\theta(t - jh) \right]. \qquad (30)$$

It follows from (30), that derivatives of the signal $w$ should be bounded up to the $(\gamma - r - 1)$th order for $r \leq \gamma$. Taking into account (23) and (30), transform equation (24) to the form

$$\dot{\varepsilon}(t) = A\varepsilon(t) - \alpha kB_1 \left[ \sum_{i=1}^{\gamma-1} \frac{d_i}{h^i} \sum_{j=1}^{i} (-1)^j C_i^j J\theta(t - jh) \right] \\ + \alpha kB_1 \left( \sum_{j=r}^{\gamma-1} \rho_{1,j} J \left( M^j \theta + G_j \varepsilon + \sum_{i=0}^{j-1} M^{j-i-1} Nw^{(i)} \right) \right. \\ \left. + \rho_2^T \theta(t) - \sum_{i=0}^{\gamma-1} \frac{d_i}{h^i} J\theta(t) \right) + \alpha kB_2 J\varepsilon(t) - \alpha kB_2 J\theta(t) \\ + B_3 f(t). \qquad (31)$$

Denote

$$\varepsilon_p(t) = \left[ \varepsilon^T(t), \theta^T(t) \right]^T,$$

$$A_p = \begin{bmatrix} A + \alpha kB_1 \sum_{j=r}^{\gamma-1} \rho_{1,j} JG_j + \alpha kB_2 J & NJ \\ \alpha kB_1 \left( \sum_{j=r}^{\gamma-1} \rho_{1,j} JM^j + \rho_2^T - \sum_{i=0}^{\gamma-1} \frac{d_i}{h^i} J \right) - \alpha kB_2 J & M \end{bmatrix},$$

$$F_{ij} = \begin{bmatrix} 0 & (-1)^{j+1} \alpha kB_1 \frac{d_i}{h^i} C_i^j J \\ 0 & 0 \end{bmatrix}, \quad i = \overline{1, \gamma-1}, \quad j = \overline{1, i},$$

$$\vartheta = \begin{bmatrix} 0 \\ N \end{bmatrix} w + \sum_{j=r}^{\gamma-1} \sum_{i=0}^{j-1} \begin{bmatrix} \alpha k\rho_{1,j} B_1 JM^{j-i-1}N \\ 0 \end{bmatrix} w^{(i)} + \begin{bmatrix} B_3 \\ 0 \end{bmatrix} f.$$

Here $\vartheta$ is a bounded signal. Taking into account the notations, rewrite systems (26) and (31) in the following form

$$\dot{\varepsilon}_p(t) = A_p \varepsilon_p(t) + \sum_{i=1}^{\gamma-1} \sum_{j=1}^{i} F_{ij} \varepsilon_p(t - jh) + \vartheta(t), \qquad (32)$$

Consider Lyapunov-Krasovskii functional

$$V = \varepsilon_p^T(t) P \varepsilon_p(t) + \sum_{i=1}^{\gamma-1} \sum_{j=1}^{i} \int_{-jh}^{0} \varepsilon_p^T(t+s) N_{ij} \varepsilon_p(t+s) ds, \qquad (33)$$

where $P = P^T > 0$ is the solution of $A_p^T P + PA_p = -Q$, $Q = Q^T > 0$, $N_{ij} = N_{ij}^T > 0$. Taking the derivative of (33) along trajectory of system (32), we get

$$\dot{V} = -\varepsilon_p^T(t) Q \varepsilon_p(t) + 2\varepsilon_p^T(t) P \sum_{i=1}^{\gamma-1} \sum_{j=1}^{i} F_{ij} \varepsilon_p(t - jh) \\ + \sum_{i=1}^{\gamma-1} \sum_{j=1}^{i} \left( \varepsilon_p^T(t) N_{ij} \varepsilon_p(t) - \varepsilon_p^T(t - jh) N_{ij} \varepsilon_p(t - jh) \right) \qquad (34) \\ + 2\varepsilon_p^T(t) P \vartheta(t).$$

Consider upper bounds of terms in (34)

$$2\varepsilon_p^T(t) P \sum_{i=1}^{\gamma-1} \sum_{j=1}^{i} F_{ij} \varepsilon_p(t - jh) \leq 0.5\gamma(\gamma-1)\chi \varepsilon_p^T(t) P^2 \varepsilon_p(t) \\ + \chi^{-1} \sum_{i=1}^{\gamma-1} \sum_{j=1}^{i} \varepsilon_p^T(t - jh) F_{ij}^T F_{ij} \varepsilon_p(t - jh),$$

$$2\varepsilon_p^T(t) P\vartheta(t) \leq \chi \varepsilon_p^T(t) P^2 \varepsilon_p(t) + \chi^{-1} |\vartheta(t)|^2,$$

where $\chi > 0$.

Taking into account upper bounds, rewrite (34) in the following form

$$\dot{V} \leq -\varepsilon_p^T(t) W \varepsilon_p(t) - \sum_{i=1}^{\gamma-1} \sum_{j=1}^{i} \varepsilon_p^T(t - jh) R_{ij} \varepsilon_p(t - jh) + \chi^{-1}\tau, \qquad (35)$$

where

$$W = Q - 0.5\gamma(\gamma-1)\chi P^2 - \chi P^2 - \sum_{i=1}^{\gamma-1} \sum_{j=1}^{i} N_{ij},$$

$$R_{ij} = N_{ij} - \chi^{-1} F_{ij}^T F_{ij}, \quad \tau = \sup_t |\vartheta(t)|^2.$$

Obviously, there exist coefficients α and χ such that $W > 0$ and $R_{ij} > 0$. We get upper bound of (35) in the form

$$\dot{V} \leq -\lambda_{\min}(W)\varepsilon_p^T(t)\varepsilon_p(t) + \chi^{-1}\tau. \quad (36)$$

Here $\lambda_{\min}(W)$ is the smallest eigenvalue of matrix $W$. It follows from (36), that $|z(t)| \leq |\varepsilon_p(t)| \leq \sqrt{\chi^{-1}\tau/\lambda_{\min}(W)}$, therefore, $\delta_1 \leq \sqrt{\chi^{-1}\tau/\lambda_{\min}(W)}$.

2) Let $r \geq \gamma$. Rewrite the operator $D(p)$ in the following form

$$D(p) = \rho^T [p^{\gamma-1}, \; p^{\gamma-2}, \; \ldots, \; 1].$$

Taking into account (33), rewrite the signal $g(t)$ as follows

$$g(t) = \rho^T \theta(t) - \sum_{i=0}^{\gamma-1} \frac{d_i}{h^i} J\theta(t) - \sum_{i=1}^{\gamma-1}\left[\frac{d_i}{h^i}\sum_{j=1}^{i}(-1)^j C_i^j J\theta(t-jh)\right]. \quad (37)$$

Substituting (37) into (24), transform equation (24) to the form

$$\dot{\varepsilon}(t) = A\varepsilon(t) + \alpha k B_1\left(\rho^T \theta(t) - \sum_{i=0}^{\gamma-1}\frac{d_i}{h^i}J\theta(t)\right)$$
$$- \alpha k B_1\left[\sum_{i=1}^{\gamma-1}\frac{d_i}{h^i}\sum_{j=1}^{i}(-1)^j C_i^j J\theta(t-jh)\right] + \alpha k B_2 J\varepsilon(t) \quad (38)$$
$$- B_2 J\theta(t) + B_3 f(t).$$

Rewrite systems (26) and (38):

$$\dot{\varepsilon}_p(t) = A_p \varepsilon_p(t) + \sum_{i=1}^{\gamma-1}\sum_{j=1}^{i} F_{ij}\varepsilon_p(t-jh) + \vartheta(t), \quad (39)$$

where
$$A_p = \begin{bmatrix} A + B_2 J & \alpha k B_1\left(\rho^T - \sum_{i=0}^{\gamma-1}\frac{d_i}{h^i}J\right) - \alpha k B_2 J \\ NJ & M \end{bmatrix},$$

$\vartheta = \begin{bmatrix} B_3 \\ 0 \end{bmatrix} f$ is a bounded function, matrix $F_{ij}$ structure corresponds to matrix $F_{ij}$ one in (32).

Since system (39) structure is similar to system (32) one, than further proof of the second case is similar to the first one. Theorem 3 is proved.

## 5. EXAMPLE

Consider the plant in the following form

$$\begin{aligned}(p^4 + q_3 p^3 + q_2 p^2 + q_1 p + q_0)z(t) &= u(t) + f(t), \\ y(t) &= z(t) + w(t). \end{aligned} \quad (40)$$

The set $\Xi$ of parameters possible values in (40) is given by inequalities:

$-1 \leq q_3 \leq 0.1, -2 \leq q_2 \leq 2, -3 \leq q_1 \leq 3, -1 \leq q_0 \leq 1$.
Additionally, $|f(t)| \leq 1$.

We choose $\sigma_i = 1$ and $\mu = 0.01$ in (3). The parameter $r$ in (3) will be determined in Table 1.

Let $\alpha = 7$, $d_0 = 0.9$, $d_1 = 1.5$, $d_2 = 2$ and $d_3 = 0.5$. Then control law (23) could be rewritten in the following form

$$u(t) = -7\left(0.9\bar{y}(t) + 1.5\bar{y}^{(1)}(t) + 2\bar{y}^{(2)}(t) + 0.5\bar{y}^{(3)}(t)\right). \quad (41)$$

We use observer (22) for estimation of derivatives in (41). Let $h = 1/20$. Then the observer (22) is rewritten in the form

$$\begin{aligned}\bar{y}(t) &= \hat{y}(t), \\ \bar{y}^{(1)}(t) &= 20[\bar{y}(t) - \bar{y}(t-0.05)], \\ \bar{y}^{(2)}(t) &= 20[\bar{y}^{(1)}(t) - \bar{y}^{(1)}(t-0.05)], \\ \bar{y}^{(3)}(t) &= 20[\bar{y}^{(2)}(t) - \bar{y}^{(2)}(t-0.05)].\end{aligned} \quad (42)$$

In addition, compare algorithm (3), (41), (42) with the classical high-gain observer (Esfandiary and Khalil, 1992) and modified high-gain observer (Astolfi and Marconi, 2015; Wang et al., 2015). The control laws (41) are the same in all algorithms.

1) Introduce high-gain observer (Esfandiary and Khalil, 1992):

$$\begin{aligned}\dot{\xi}(t) &= \begin{bmatrix} 0 & 1 & 0 & 0 \\ 0 & 0 & 1 & 0 \\ 0 & 0 & 0 & 1 \\ 0 & 0 & 0 & 0 \end{bmatrix}\xi(t) + \begin{bmatrix} 110 \\ 110^2 \cdot 0.35 \\ 110^3 \cdot 0.05 \\ 110^4 \cdot 0.0024 \end{bmatrix} \\ &\quad \times (y(t) - [1 \; 0 \; 0 \; 0]\xi(t)), \end{aligned} \quad (43)$$

$\bar{y}(t) = [1 \; 0 \; 0 \; 0]\xi(t), \; \bar{y}^{(1)}(t) = [0 \; 1 \; 0 \; 0]\xi(t),$
$\bar{y}^{(2)}(t) = [0 \; 0 \; 1 \; 0]\xi(t), \; \bar{y}^{(3)}(t) = [0 \; 0 \; 0 \; 1]\xi(t);$

2) Consider modified high-gain observer (Astolfi and Marconi, 2015; Wang et al., 2015):

$$\begin{aligned}\dot{\eta}_1(t) &= \begin{bmatrix} 0 & 1 \\ 0 & 0 \end{bmatrix}\eta_1(t) + \begin{bmatrix} 0 & 0 \\ 0 & 1 \end{bmatrix}\eta_2(t) \\ &\quad + \begin{bmatrix} 110 & 0 \\ 0 & 110^2 \end{bmatrix}\begin{bmatrix} 0.5 \\ 0.16 \end{bmatrix}(y(t) - [1 \; 0]\eta_1(t)), \\ \dot{\eta}_2(t) &= \begin{bmatrix} 0 & 1 \\ 0 & 0 \end{bmatrix}\eta_2(t) + \begin{bmatrix} 0 & 0 \\ 0 & 1 \end{bmatrix}\eta_3(t) \\ &\quad + \begin{bmatrix} 110 & 0 \\ 0 & 110^2 \end{bmatrix}\begin{bmatrix} 0.5 \\ 0.0525 \end{bmatrix}([0 \; 1]\eta_1(t) - [1 \; 0]\eta_2(t)), \\ \dot{\eta}_3(t) &= \begin{bmatrix} 0 & 1 \\ 0 & 0 \end{bmatrix}\eta_3(t) \\ &\quad + \begin{bmatrix} 110 & 0 \\ 0 & 110^2 \end{bmatrix}\begin{bmatrix} 0.5 \\ 0.0171 \end{bmatrix}([0 \; 1]\eta_2(t) - [1 \; 0]\eta_3(t)), \end{aligned} \quad (44)$$

$\bar{y}(t) = [1 \; 0]\eta_1(t), \; \bar{y}^{(1)}(t) = [1 \; 0]\eta_2(t),$
$\bar{y}^{(2)}(t) = [1 \; 0]\eta_3(t), \; \bar{y}^{(3)}(t) = [0 \; 1]\eta_3(t).$

Let $q_3 = 0$, $q_2 = 1$, $q_1 = 1$, $q_0 = 0$, $f(t) = 0$, $w(t) = \sin(0.5 \cdot 10^3 t)$ and $z(0) = 1$, $\dot{z}(0) = \ddot{z}(0) = 0$, $\dddot{z}(0) = -1$ in (40). Table 1 shows the maximum errors of estimation of the signal $z(t)$ derivatives at steady-state mode using control system (3), (41), (42), control system (41), (43) and control system (41), (44).

**Table 1**. The value of $e^{(i)}(t) = \sup_t |z^{(i)}(t) - \overline{y}^{(i)}(t)|$, $i = \overline{0,3}$ at steady-state mode (after 8 s) for the proposed algorithm, algorithms (43) and algorithm (44)

| Control system | $e(t)$ | $e^{(1)}(t)$ | $e^{(2)}(t)$ | $e^{(3)}(t)$ |
|---|---|---|---|---|
| Control system (41), (43) (high-gain observer) | 0.22 | 8.41 | 132.3 | 698 |
| Control system (41), (44) (modified high-gain observer) | 0.26 | 4.9 | 31.1 | 266.1 |
| Proposed control system (3), (41), (42) when $r = 2$ in (3) | 0.04 | 0.2 | 2 | 57 |
| Proposed control system (3), (41), (42) when $r = 5$ in (3) | $3 \cdot 10^{-3}$ | $7.5 \cdot 10^{-4}$ | $2 \cdot 10^{-3}$ | $5 \cdot 10^{-3}$ |

Table 1 shows that the proposed control algorithm can significantly reduce the estimate error of derivatives of signal $z$. However, the dynamical order of the proposed algorithm for $r = 2$ is one less than the dynamical order of the algorithm (44). Furthermore, it follows from Table 1 that increasing the parameter $r$ can improve the quality of derivative estimates.

Let $q_3 = 0.1$, $q_2 = 2$, $q_1 = 3$, $q_0 = 1$, $f(t) = \sin t$, $w(t) = \sin(0.5 \cdot 10^3 t) + \sin(10^3 t) + \sin(10^4 t)$, $z(0) = 1$, $\dot{z}(0) = \ddot{z}(0) = 0$, $\dddot{z}(0) = -1$ in (40). Fig. 1 shows the simulation result of $z(t)$, $\dot{z}(t)$, $\ddot{z}(t)$ и $\dddot{z}(t)$ using the proposed control algorithm (3), (41), (42) for $r = 5$ in (3). The simulation results of $z(t)$ and $\ddot{z}(t)$ are represented by continuous curves and the simulation results of $\dot{z}(t)$ and $\dddot{z}(t)$ are represented by dashed ones.

The simulation results show (Fig. 1) that after 10 (s) the absolute values of the signals $z(t)$, $\dot{z}(t)$, $\ddot{z}(t)$ and $\dddot{z}(t)$ do not exceed 0.014. However, the absolute values of estimate errors of $z(t)$, $\dot{z}(t)$, $\ddot{z}(t)$ and $\dddot{z}(t)$ do not exceed $2 \cdot 10^{-3}$.

## 6. CONCLUSION

In this paper the robust control algorithm under parametric uncertainties, external bounded disturbances and high-frequency noises in measurement signal was proposed. For synthesis of control algorithm we used the approach that allows one to control independently the quality of noise filtering and the quality of the error of stabilization of the output variable. The simulation results show the effectiveness of the proposed algorithm as compared as standard high-gain observer (Esfandiary and Khalil, 1992) and modified high-gain observer (Astolfi and Marconi, 2015; Wang et al., 2015).

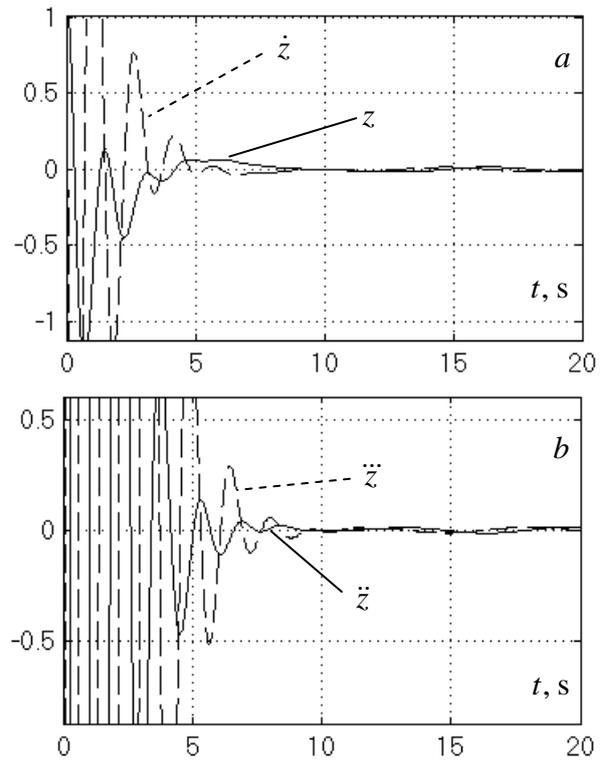

Fig. 1. The simulation results $z(t)$, $\dot{z}(t)$ (Fig. 1, a) and $\ddot{z}(t)$, $\dddot{z}(t)$ (Fig. 1, b).